# Comparison between theoretical four-loop predictions and Monte Carlo calculations in the two-dimensional $N$-vector model for $N = 3, 4, 8$

Sergio Caracciolo[a,*], Robert G. Edwards[b], Tereza Mendes[c], Andrea Pelissetto[d] and Alan D. Sokal[c]

[a]Dipartimento di Fisica and INFN, Università degli Studi di Lecce, Lecce 73100, ITALIA

[b]SCRI, Florida State University, Tallahassee, FL 32306 USA

[c]Department of Physics, New York University, 4 Washington Place, New York, NY 10003 USA

[d]Dipartimento di Fisica and INFN, Università degli Studi di Pisa, Pisa 56100, ITALIA

We have computed the four-loop contribution to the beta-function and to the anomalous dimension of the field for the two-dimensional lattice $N$-vector model. This allows the determination of the second perturbative correction to various long-distance quantities like the correlation lengths and the susceptibilities. We compare these predictions with new Monte Carlo data for $N = 3, 4, 8$. From these data we also extract the values of various universal nonperturbative constants, which we compare with the predictions of the $1/N$ expansion.

The $N$-vector model describes configurations of classical spins taking values on the unit sphere $S^{N-1} \subset \mathbb{R}^N$. We consider here the standard nearest-neighbor action

$$\mathcal{H}(\boldsymbol{\sigma}) \;=\; -\beta \sum_{\langle xy \rangle} \boldsymbol{\sigma}_x \cdot \boldsymbol{\sigma}_y \; . \tag{1}$$

The perturbative renormalization group predicts that when $\beta \nearrow \infty$, the (infinite-volume) long-distance quantities of lattice theory should give the same results as the continuum theory in the $\overline{\rm MS}$ normalization, provided that one rescales lengths by the factor

$$\Lambda \;=\; e^{-2\pi\beta/(N-2)} \left(\frac{2\pi\beta}{N-2}\right)^{\frac{1}{N-2}} 2^{5/2} e^{\frac{\pi}{2(N-2)}} \; . \tag{2}$$

For the isovector and isotensor two-point functions

$$G_V(x,y) \;=\; \langle \boldsymbol{\sigma}_x \cdot \sigma_y \rangle \tag{3}$$

$$G_T(x,y) \;=\; \langle (\boldsymbol{\sigma}_x \cdot \sigma_y)^2 \rangle - \frac{1}{N} \tag{4}$$

we shall consider the RG predictions for the correlation lengths

$$\xi_\#(\beta) \;=\; \widetilde{C}_{\xi_\#} \Lambda^{-1} \left[1 + \sum_{i=1}^{\infty} \frac{a_i}{\beta^i}\right] \tag{5}$$

and the susceptibilities

$$\chi_V(\beta) = \widetilde{C}_{\chi_V} \Lambda^{-2} \left(\frac{2\pi\beta}{N-2}\right)^{-\frac{N-1}{N-2}} \left[1 + \sum_{i=1}^{\infty} \frac{b_i}{\beta^i}\right] \tag{6}$$

$$\chi_T(\beta) = \widetilde{C}_{\chi_T} \Lambda^{-2} \left(\frac{2\pi\beta}{N-2}\right)^{-\frac{2N}{N-2}} \left[1 + \sum_{i=1}^{\infty} \frac{d_i}{\beta^i}\right] \tag{7}$$

Here the constants $\widetilde{C}_\#$ are *universal* but nonperturbative, while the coefficients $a_i$, $b_i$ and $d_i$ can be determined at the $(i+2)$-th order of the perturbative expansion. There is a prediction [1] for the exponential correlation length (= inverse mass gap) in the isovector channel:

$$\widetilde{C}_{\xi_V^{(exp)}} \;=\; \left(\frac{e}{8}\right)^{1/(N-2)} \Gamma\left(1 + \frac{1}{N-2}\right) . \tag{8}$$

As the model should not have bound states, in the isotensor channel one expects

$$\widetilde{C}_{\xi_T^{(exp)}} \;=\; \frac{1}{2} \widetilde{C}_{\xi_V^{(exp)}} \; . \tag{9}$$

The constants related to the second-moment correlation lengths and the susceptibilities are known analytically only in the large-$N$ expansion,

*Speaker at the conference.

namely, through order $1/N$ both in the isovector [2] and isotensor [3] sectors. In particular,

$$\frac{\widetilde{C}_{\xi_V^{(2)}}}{\widetilde{C}_{\xi_V^{(exp)}}} \; = \; 1 - \frac{0.003225}{N} + O(1/N^2) \; , \qquad (10)$$

so that even for $N = 3$ this ratio differs only marginally from 1 (in good agreement with Monte Carlo simulations [4]).

We can also try an *improved expansion parameter* [5,6] based on the isovector energy $E_V = \langle \boldsymbol{\sigma}_{0,0} \cdot \boldsymbol{\sigma}_{1,0} \rangle$: we define

$$\beta_{\mathit{eff}} \; \equiv \; \frac{N-1}{4(1-E_V)} \; = \; \beta + O(1) \; . \qquad (11)$$

This $\beta_{\mathit{eff}}$ has the property that at $N \nearrow \infty$ with $\widetilde{\beta} \equiv \beta/N$ fixed, there are only exponentially small corrections to the two-loop predictions for correlation lengths and susceptibilities.

A precise confirmation from Monte Carlo simulations of (5) *with the correct nonperturbative constant* (8)/(10) would be good evidence in favor of the conventional asymptotic-freedom picture [7], which has been criticized [8]. At the last Lattice conference [9] we presented data for the case $N = 3$ at infinite-volume correlation lengths $\xi_\infty$ up to $\approx 10^5$ [10], obtained by using finite-size-scaling extrapolation at fixed $\beta$ [11]. The discrepancy of these data from the three-loop predictions [12,6] was already quite small ($\sim 4\%$), and in the "improved expansion parameter" the agreement was even better. For this reason we considered it worthwhile to compute the next-order perturbative correction, in order to see whether the remaining discrepancy — which is nevertheless larger than the estimated statistical error — can be removed. We therefore computed the four-loop contributions to the beta-function and the anomalous dimension of the field for the lattice model (1); from these we derived the $1/\beta^2$ corrections to the correlation lengths and the general spin-$n$ susceptibility [13]. For example, we have

$$a_2 \; = \; \frac{0.0688 - 0.0028N + 0.0107N^2 - 0.0129N^3}{(N-2)^2} \; . \qquad (12)$$

In Figure 1 we plot $\xi^{(2)}_{\infty,estimate}$ divided by the two-loop, three-loop and four-loop predictions

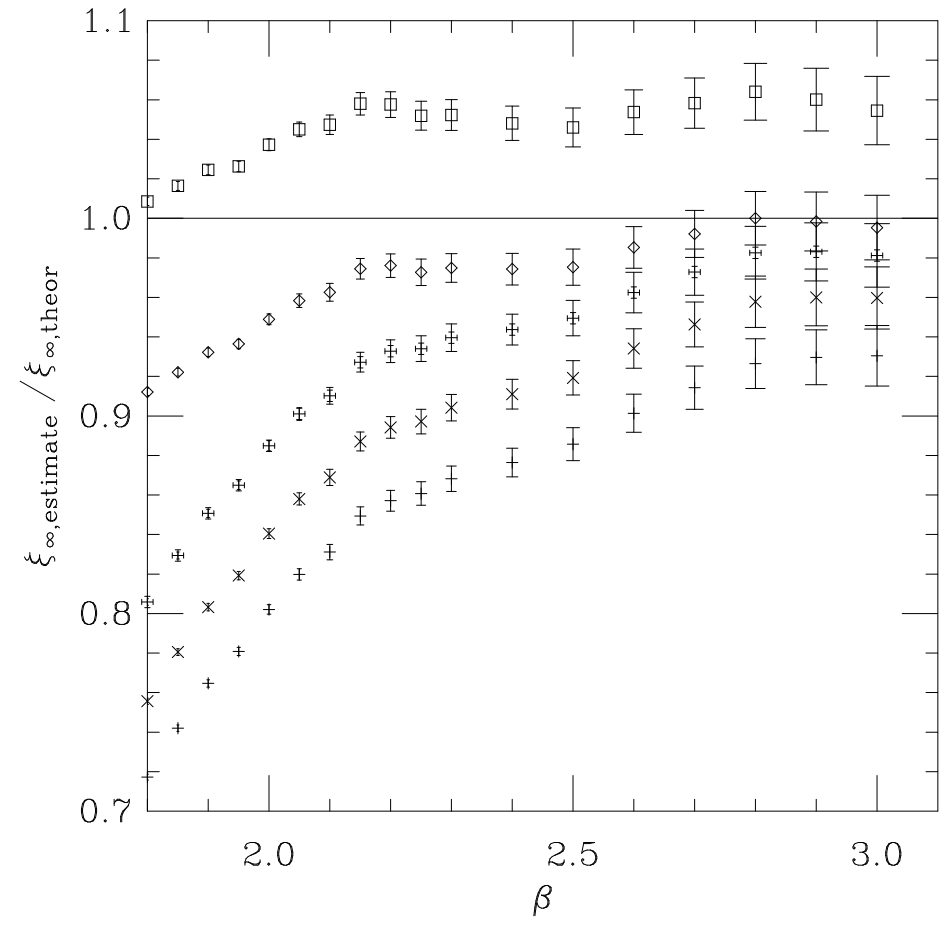

Figure 1. $\xi^{(2)}_{\infty,estimate}/\xi^{(2)}_{\infty,theor}$ versus $\beta$ for the $O(3)$ model. Error bars are one standard deviation (statistical error only). There are five versions of $\xi^{(2)}_{\infty,theor}$: standard perturbation theory in $1/\beta$ gives points $+$ (2-loop), $\times$ (3-loop) and ⧺ (4-loop); "improved" perturbation theory in $1/\beta_{\mathit{eff}}$ gives points $\Box$ (2-loop) and $\Diamond$ (3-loop).

(points $+$, $\times$ and ⧺) or by the "improved" two-loop and three-loop predictions (points $\Box$ and $\Diamond$). The four-loop truncation of (5) is now fully compatible with our last extrapolated point. It falls roughly halfway between the three-loop truncations in $1/\beta$ and $1/\beta_{\mathit{eff}}$.

A similar result is obtained for $N = 4$, where the central estimate of our last point (which is at the rather small correlation length $\xi_\infty \approx 155$) differs from the theoretical four-loop prediction by only 4–5%; and for $N = 8$ [14], where the central estimate of our last point (which is at $\xi_\infty \approx 650$) shows an extremely good agreement (better than 1%).

We have also tried to take into account higher-loop corrections by using information from the $1/N$-expansion. Let

$$a_n^{(1/N)}(N) \; = \; N^{n-1}\overline{a}_n \qquad (13)$$

be the leading contribution to $a_n$ in the limit $N \nearrow \infty$. We computed $\overline{a}_n$ up to $n = 8$ [13]. How good is the approximation in which only such a term is retained? We can compare with the known coefficients $a_1$ and $a_2$. For $N = 4$, we have

$$\frac{a_1(4)}{a_1^{(1/N)}(4)} = 3.73; \qquad \frac{a_2(4)}{a_2^{(1/N)}(4)} = 2.88 \; ; \tag{14}$$

while for $N = 8$, we have

$$\frac{a_1(8)}{a_1^{(1/N)}(8)} = 1.91; \qquad \frac{a_2(8)}{a_2^{(1/N)}(8)} = 1.58 \; . \tag{15}$$

The convergence seems slow; indeed, only at $N \gtrsim 50$ are the first two coefficients correct to within 10%. Let us now define $k_n$ by

$$a_n(8) \;=\; k_n 8^{n-1} \overline{a}_n \; . \tag{16}$$

Already at $\beta = 5.80$ ($\xi_\infty \approx 33$) the Monte Carlo value [14,15] is in good agreement with the theoretical predictions. Indeed we get

$$\frac{\xi_{MC}}{\xi_{th}^{8-loop}} \;=\; \begin{cases} 0.998 & \text{when } k_n = 1 \;\forall n \\ 1.001 & \text{when } k_n = 2 \;\forall n \end{cases} \tag{17}$$

with a statistical error of $\pm 0.002$. For larger values of $\beta$ this ratio remains roughly constant, although the error bars grow. We can thus claim a nice control of (5) for $N = 8$.

Having verified (5), we can now extract from the Monte Carlo data a numerical evaluation of the nonperturbative universal constants $\widetilde{C}_\#$. For the limiting ratio $\xi_V^{(2)}/\xi_T^{(2)}$ we have the numerical results 3.51 for $N = 3$, 3.14 for $N = 4$, and 2.77 for $N = 8$, with error bars less than $\pm 0.01$, which can be compared with the $1/N$-expansion prediction [3]

$$\frac{\widetilde{C}_{\xi_V^{(2)}}}{\widetilde{C}_{\xi_T^{(2)}}} \;=\; \sqrt{6}\left[1 + \frac{1.1999}{N} + \ldots\right] \tag{18}$$

$$\approx \; \begin{cases} 3.43 & \text{for } N = 3 \\ 3.18 & \text{for } N = 4 \\ 2.82 & \text{for } N = 8 \end{cases} \tag{19}$$

To extrapolate the asymptotic value of the isovector susceptibility, we found it convenient to study the dimensionless ratio

$$\frac{\chi_V}{(\xi_V^{(2)})^2} \;=\; \frac{\widetilde{C}_{\chi_V}}{\widetilde{C}^2_{\xi_V^{(2)}}}\left(\frac{2\pi\beta}{N-2}\right)^{-\frac{N-1}{N-2}}\left[1 + \sum_{i=1}^{\infty}\frac{c_i}{\beta^i}\right] \tag{20}$$

because our knowledge of the lattice beta and gamma functions at four loops allows the determination of the first *three* coefficients $c_i$ [6,13]. We

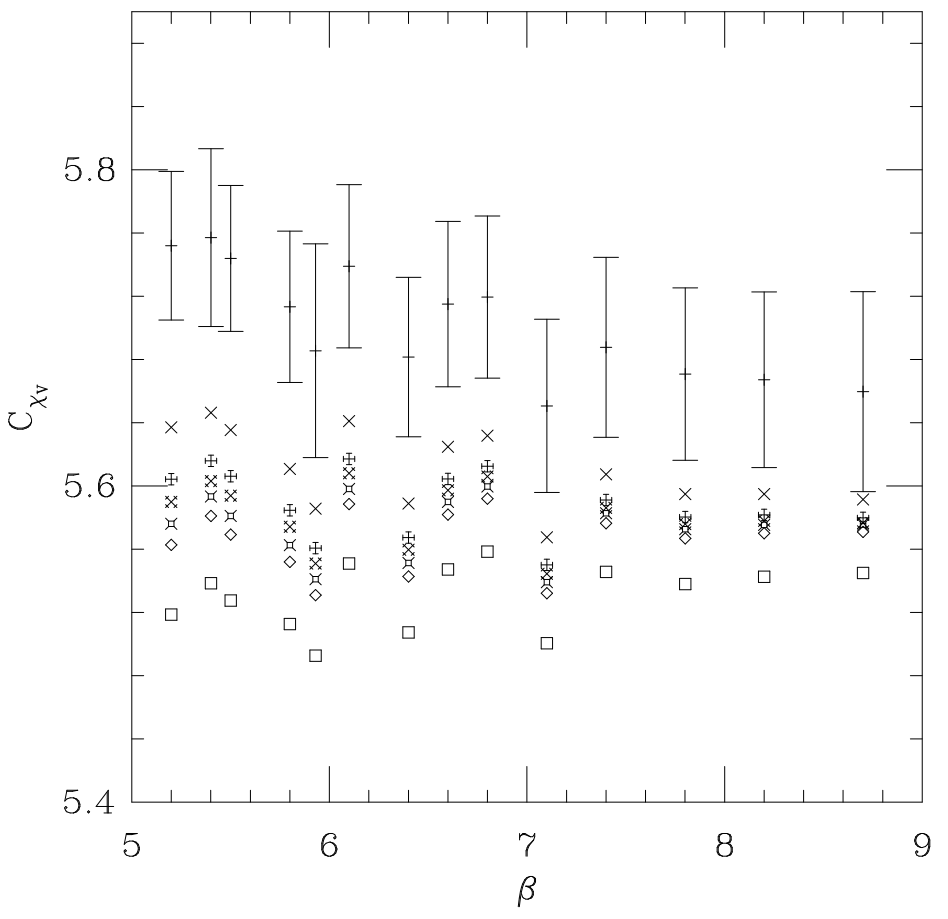

Figure 2. Estimate of $\widetilde{C}_{\chi_V}$ [from (20)/(8)/(10)] versus $\beta$ for the $O(8)$ model. Error bars (one standard deviation, statistical error only) shown for clarity only on one set of points. There are seven versions of $\chi_V/(\xi_V^{(2)})^2$: standard perturbation theory in $1/\beta$ gives points + (leading), × (with $c_1$), ⊞ ($c_{1,2}$) and ⋇ ($c_{1,2,3}$); "improved" perturbation theory in $1/\beta_{\mathit{eff}}$ gives points □ (leading), ◇ ($c'_1$) and ⌘ ($c'_{1,2}$).

then use the exact formula (8)/(10) to estimate $\widetilde{C}_{\chi_V}$: see Figure 2 for $N = 8$. We find

$$\widetilde{C}_{\chi_V} \;=\; \begin{cases} 10.8 \pm 0.8 & \text{for } N = 3 \\ 5.9 \pm 0.1 & \text{for } N = 4 \\ 5.6 \pm 0.1 & \text{for } N = 8 \end{cases} \tag{21}$$

which can be compared with the $1/N$-expression [2]

$$\widetilde{C}_{\chi_V} \;=\; 2\pi\left[1 + \frac{4 + 3\gamma_C - 3\gamma_E - 7\log 2}{N}\right] \tag{22}$$

$$\approx \; \begin{cases} 3.67 & \text{for } N = 3 \\ 4.32 & \text{for } N = 4 \\ 5.30 & \text{for } N = 8 \end{cases} \tag{23}$$

where $\gamma_E$ is Euler's constant and

$$\gamma_C \;=\; \log\frac{\Gamma(1/3)\Gamma(7/6)}{\Gamma(2/3)\Gamma(5/6)} \; . \tag{24}$$

Clearly the $O(1/N^2)$ corrections are significant! We have made the *wild guess* that the exact ex-

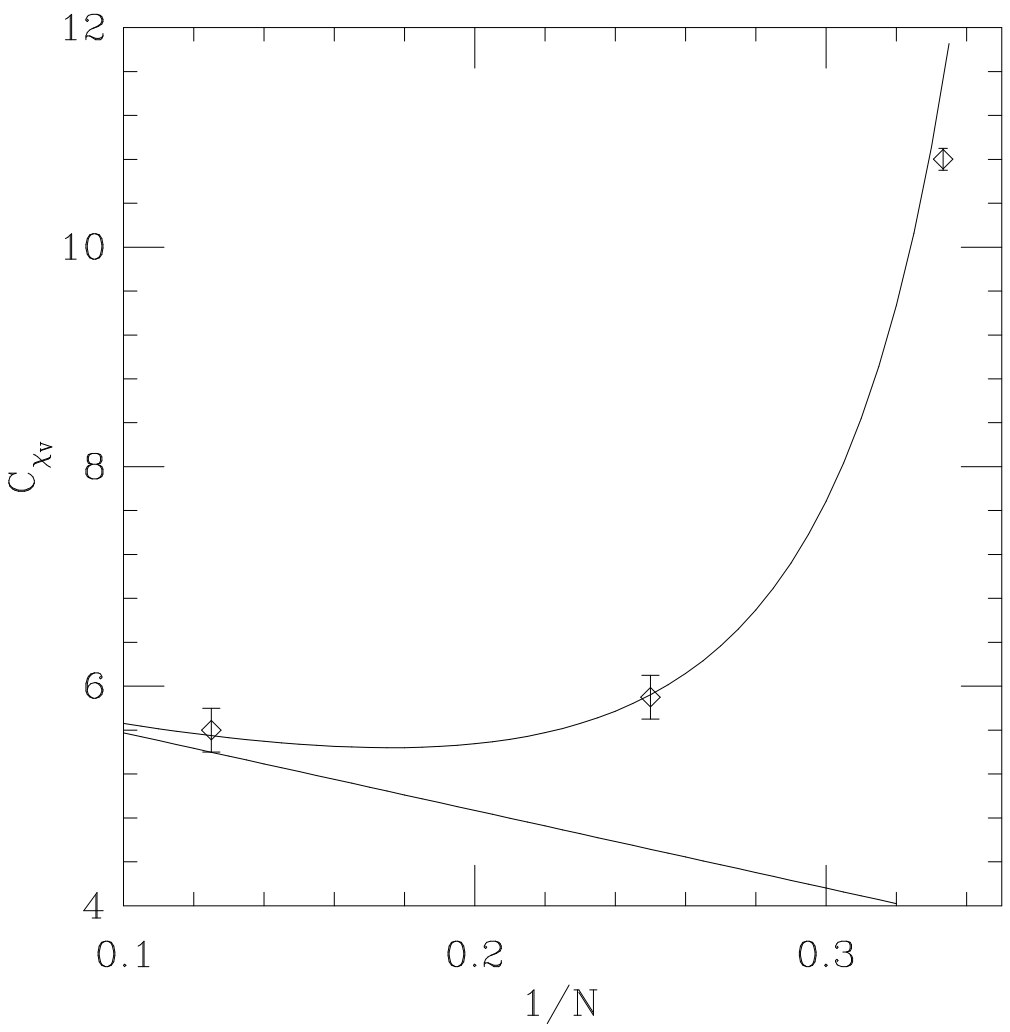

Figure 3. $\widetilde{C}_{\chi_V}$ versus $1/N$. The straight line is the $1/N$-prediction (22), while the curve is the wild guess (25); the points $\diamond$ are our Monte Carlo estimates.

pression for $\widetilde{C}_{\chi_V}$ is

$$\widetilde{C}_{\chi_V} \;=\; 2\pi \left(\frac{e^{4+3\gamma_C}}{128}\right)^{\frac{1}{N-2}} \Gamma^3\!\left(1+\frac{1}{N-2}\right) \;, \quad (25)$$

in analogy with (8). In Figure 3 we compare our Monte Carlo results with the $1/N$-prediction (22) and the wild guess (25). The result for $N = 3$ is close to the guessed formula but outside the statistical errors, while the values for $N = 4, 8$ follow the Ansatz nicely. It would be interesting to test (25) by computing the $O(1/N^2)$ term in (22).

The agreement with the $1/N$-expansion is much poorer for the isotensor susceptibility: our Monte Carlo data yield

$$\widetilde{C}_{\chi_T} \;=\; \begin{cases} 1200 \pm 100 & \text{for } N = 3 \\ 23 \pm 2 & \text{for } N = 4 \\ 4.7 \pm 0.2 & \text{for } N = 8 \end{cases} \quad (26)$$

compared to the $1/N$-expansion result [3]

$$\widetilde{C}_{\chi_T} \;=\; \pi \left[1 - \frac{0.0296}{N} + \ldots\right] . \quad (27)$$

Clearly the $O(1/N^2)$ and higher corrections must have a *drastic* effect for $N \lesssim 20$.

The authors' research was supported by CNR, INFN, DOE contracts DE-FG05-85ER250000 and DE-FG05-92ER40742, NSF grant DMS-9200719, and NATO CRG 910251.